\documentclass[11pt,a4pape1r]{article}
\usepackage{jheppub}

\usepackage{amssymb}
\usepackage{amsmath}
\usepackage{amsfonts}

\usepackage{ctable}
\usepackage{amsmath}
\usepackage{amsfonts}
\newcommand{\TCM}[1]{\textcolor{black}{#1}}

\newcommand{\mumu}{\ensuremath{\mu^{+}\mu^{-}}}

\newcommand{\B}{\ensuremath{\mathcal{B}}}

\title{\boldmath Measurement of the branching fraction of $J/\psi\rightarrow\rho\pi$ at KEDR}

\author[a]{V.V.~Anashin,}
\author[a]{O.V.~Anchugov,}
\author[a]{A.V.~Andrianov,}
\author[a]{K.V.~Astrelina,}
\author[a,b]{V.M.~Aulchenko,}
\author[a,b]{E.M.~Baldin,}
\author[a,c]{G.N.~Baranov,} 
\author[a]{A.K.~Barladyan,}

\author[a,b,c]{A.Yu.~Barnyakov,}
\author[a,b,c]{M.Yu.~Barnyakov,}
\author[a]{I.Yu.~Basok,}
\author[a]{A.M.~Batrakov,}
\author[a]{E.A.~Bekhtenev,}
\author[a]{O.V.~Belikov,}
\author[a]{D.E.~Berkaev,}
\author[a,b]{A.E.~Blinov,}

\author[a,b,c]{V.E.~Blinov,}
\author[a]{M.F.~Blinov,}
\author[a,b]{A.V.~Bobrov,}
\author[a,b]{V.S.~Bobrovnikov,}
\author[a,b]{A.V.~Bogomyagkov,}
\author[a]{D.Yu.~Bolkhovityanov,}
\author[a,b]{A.E.~Bondar,}

\author[a,b]{A.R.~Buzykaev,}
\author[a,b]{P.B.~Cheblakov,}
\author[a,c]{V.L.~Dorohov,}
\author[a]{F.A.~Emanov,}
\author[a]{V.V.~Gambaryan,}

\author[a,b,c]{D.N.~Grigoriev,}
\author[a]{V.V.~Kaminskiy,}
\author[a]{S.E.~Karnaev,}
\author[a]{G.V.~Karpov,}
\author[a]{S.V.~Karpov,}
\author[a,c]{K.Yu.~Karukina,}
\author[a]{D.P.~Kashtankin,}
\author[a]{P.V.~Kasyanenko,}
\author[a]{A.A.~Katcin,}

\author[a]{T.A.~Kharlamova,}
\author[a]{V.A.~Kiselev,}
\author[a,b]{S.A.~Kononov,}
\author[a]{A.A.~Krasnov,}
\author[a,b]{E.A.~Kravchenko,}
\author[a,b]{V.N.~Kudryavtsev,}
\author[a,b]{V.F.~Kulikov,}

\author[a]{I.A.~Kuyanov,}
\author[a,c]{E.B.~Levichev,}
\author[a]{P.V.~Logachev,}
\author[a,b]{D.A.~Maksimov,}
\author[a]{Yu.I.~Maltseva,}
\author[a]{V.M.~Malyshev,}
\author[a,b]{A.L.~Maslennikov,}

\author[a,b]{O.I.~Meshkov,}
\author[a]{S.I.~Mishnev,}
\author[a]{I.A.~Morozov,}
\author[a,b]{I.I.~Morozov,}

\author[a]{D.A.~Nikiforov,}
\author[a]{S.A.~Nikitin,}
\author[a,b]{I.B.~Nikolaev,}
\author[a]{I.N.~Okunev,}

\author[a]{S.B.~Oreshkin,}
\author[a,b]{A.A.~Osipov,}
\author[a,c]{I.V.~Ovtin,}
\author[a]{A.V.~Pavlenko,}
\author[a,b]{S.V.~Peleganchuk,}

\author[a]{P.A.~Piminov,}
\author[a]{N.A.~Podgornov,}
\author[a,b]{V.G.~Prisekin,}

\author[a,b]{O.L.~Rezanova,}
\author[a,b]{A.A.~Ruban,}
\author[a]{G.A.~Savinov,}
\author[a,b]{A.G.~Shamov,}

\author[a]{L.I.~Shekhtman,}
\author[a]{D.A.~Shvedov,}
\author[a,b]{B.A.~Shwartz,}
\author[a]{E.A.~Simonov,}

\author[a]{S.V.~Sinyatkin,}
\author[a]{A.N.~Skrinsky,}
\author[a,b]{A.V.~Sokolov,}
\author[a,b]{E.V.~Starostina,}

\author[a]{D.P.~Sukhanov,}
\author[a,b]{A.M.~Sukharev,}
\author[a,b]{A.A.~Talyshev,}
\author[a,b]{V.A.~Tayursky,}
\author[a,b]{V.I.~Telnov,}
\author[a,b]{Yu.A.~Tikhonov,}
\author[a,b,1]{K.Yu.~Todyshev,\note{Corresponding author}}
\author[a]{A.G.~Tribendis,}
\author[a]{G.M.~Tumaikin,}
\author[a]{Yu.V.~Usov,}
\author[a]{A.I.~Vorobiov,}
\author[a,b]{V.N.~Zhilich,}
\author[a]{A.A.~Zhukov,}
\author[a,b]{V.V.~Zhulanov,}
\author[a,b]{A.N.~Zhuravlev}

\affiliation[a]{Budker Institute of Nuclear Physics, SB RAS,  Novosibirsk 630090, Russia}
\affiliation[b]{Novosibirsk State University, Novosibirsk 630090, Russia}
\affiliation[c]{Novosibirsk State Technical University,  Novosibirsk 630092, Russia}
\emailAdd{todyshev@inp.nsk.su}
\date{24 november 2022}
\abstract{
We present the study of the decay $J/\psi \rightarrow \rho\pi$. 
The results are based on of 5.2~million $J/\psi$ events collected by
the KEDR detector at the VEPP-4M collider.
The branching fractions are measured to be
$\B(J/\psi\rightarrow\rho\pi)  = \big(2.072\pm 0.017 \pm 0.062 \big)\cdot 10^{-2}$
and $\B(J/\psi\rightarrow\pi^+\pi^-\pi^0)  = \big(1.878 \pm 0.013 \pm 0.051 \big)\cdot 10^{-2}$,
where the first uncertainties are statistical and the second systematic.
Our results are more precise than the previous relative measurements.
}
\begin{document}
\maketitle
\flushbottom
\section{Introduction}\label{sec:intro}
In this paper, a study for $J/\psi$ meson decays of the type
$J/\psi\rightarrow\rho\pi$ is presented. All three decay modes 
$J/\psi\rightarrow  \rho^{+}  \pi^{-}, J/\psi \rightarrow \rho^{-}\pi^{+}$ and $J/\psi\rightarrow \rho^{0}\pi^{0}$ are examined.

The probability of the  $J/\psi\rightarrow\rho\pi$ decay is
the largest among hadronic $J/\psi$ decays with an intermediate
resonance and is $\B(J/\psi\rightarrow\rho\pi)=(1.69 \pm 0.15)\%$~\cite{PDG:2022}.
Nine experiments contributed to the $J/\psi\rightarrow \rho \pi$  branching fraction measurement
\cite{MRK1:1976,CNTR:1976,DASP:1978,PLUT:1978,MRK2:1983,MRK3:1988,BES:1996,BES:2004,BABAR:2004}.
At the moment, there is a noticeable discrepancy between the results of
early experiments
\cite{MRK1:1976,CNTR:1976,DASP:1978,PLUT:1978,MRK2:1983,MRK3:1988,BES:1996}
and the latest measurements by
collaborations BES \cite{BES:2004} and BaBar \cite{BABAR:2004}. 
The Particle Data Group (PDG) gives the value of the scale factor 2.4.
This motivates us to perform  new measurement of the $J/\psi\rightarrow \rho \pi$  branching fraction.
The decay $J/\psi\rightarrow\rho\pi$ is the main process leading to the final three-pion state,
the study of which is important in itself. The branching fraction of the process
$J/\psi\rightarrow\pi^+\pi^-\pi^0$ have been measured directly by the MARKII \cite{MRK2:1983},
BES \cite{BES:2004,BES:2012} and BaBar\cite{BaBar:2007,BaBar:2021} collaborations.

In addition to the circumstances already indicated, 
the study of decays into three $\pi$ mesons is important for a better understanding 
of the rescattering effects, which, for example, are discussed in Refs. \cite{Cheng:2021,Stamen:2022}. 
Refinement of the $J/\psi\rightarrow \rho \pi$  branching fraction 
will be useful in the study of the so-called $\rho-\pi$ puzzle \cite{MRK2:1983}.
It would also be interesting to compare the value of this branching fraction 
with a relatively recent theoretical calculation given in Ref. \cite{Flores-Baez:2016}. 

We also note that the exact determination of the $J/\psi\rightarrow\rho\pi$  branching fraction
may be important in the analysis of other processes for which the decay considered 
in this article is a background process.
\section{Theoretical framework and MC simulation}
\label{sec:frame}
The differential cross section of the process $J/\psi\rightarrow \pi^+\pi^-\pi^0$ can be written as a sum of contributions
of several intermediate states $\rho(770)\pi$, $\rho(1450)\pi$, $\omega\pi$, $\rho(1700)\pi$.
In this paper, we consider the first two terms, which are dominant. 
We neglect the remaining terms  as well as contribution of the decay
$J/\psi\rightarrow \pi^+\pi^-\pi^0$ without intermediate resonances, the corresponding systematic errors are considered in Section \ref{sec:fitting}. 
Under these conditions, the expression for the differential cross section \TCM{has the form}
\begin{equation}
\frac{d\sigma}{d\Gamma} \propto \bigg|\sum\limits_{j} a_{j}+ \sum\limits_{j} b_{j}e^{i\phi}\bigg|^2=\bigg|\sum\limits_{j} a_{j}\bigg|^2+ \bigg|\sum\limits_{j} b_{j}\bigg|^2+ \sum\limits_{i,j}\bigg(a_{i} b_{j}^{*} e^{-i\phi}+a_{i}^{*}b_{j} e^{i\phi}\bigg),
\label{eq:dsdG}
\end{equation}
where ${d\Gamma}$ is a phase space element, $j$ can be $0,+,-$ corresponding to the charged states $\rho(770)$ and $\rho(1450)$. 
The amplitudes $a_{j}$ and $b_{j}$ are the functions of $s$ and pions
momenta and correspond to neutral and charged modes of $\rho(770)$ and $\rho(1450)$ resonances.
The amplitude for  neutral mode can be written as
\begin{equation}
  a_{0}=a_{\rho^0\pi^0} = ({\bf p_{+}}\times {\bf p_{-}})\sin{\theta_n}\frac{m_{\rho^0}^2}{q^2-m_{\rho^{0}}^2+i q \Gamma_{\rho^{0}}(q^2)}\sqrt{\B(J/\psi\rightarrow\rho^0\pi^0)\B(\rho^0\rightarrow\pi^+\pi^-)},
\label{eq:a0} 
\end{equation}  
  
where  {$\bf p_{+}$} and  {$\bf p_{-}$} are charged pion momenta, 
$\theta_n$ is an angle between the normal to the reaction plane and
the beam axis, $\Gamma_{\rho^{0}}(q^2)= \Gamma_{\rho^{0}}
\Big(\frac{p_{\pi}(q^2)}{p_{\pi}(m_{\rho^0}^2)}\Big)^3
  \bigg(\frac{m_{\rho^0}^2}{q^2}\bigg)$, $m_{\rho^0}$ and
  $\Gamma_{\rho^{0}}$ are the mass and the width of the
  $\rho^0(770)$, $q$ is the invariant mass of the pion pair,
  $p_\pi $ is the pion momentum in the $\rho$ rest frame. The
 amplitude  $b_{0}$ is written in the same way by replacing the
 $\rho(770)$ to $\rho(1450)$.  
 The above parametrization goes back to the work of G.J. Gounaris and J.J. Sakurai \cite{GS:1968}.

Consider, for example, one of the terms in the last sum of the
expression \eqref{eq:dsdG}
\begin{eqnarray}
\nonumber
a_{0} b_{0}^{*} e^{-i\phi}+a_{0}^{*}b_{0} e^{i\phi}  &=   \sqrt{\B(J/\psi\rightarrow\rho^0\pi^0)\B(\rho^0\rightarrow\pi^+\pi^-)} \sqrt{\B(J/\psi\rightarrow\rho^0(1450)\pi^0)\B(\rho^0(1450)\rightarrow\pi^+\pi^-)}\\ \nonumber
  & ({\bf p_{+}}\times {\bf p_{-}})^2\sin^2{\theta_n}\\ \nonumber
 &\bigg[\frac{2  m_{\rho^0}^2m_{\rho^0(1450)}^2 (q^4+m_{\rho^0}^2m_{\rho^0(1450)}^2+q^2\Gamma_{\rho^0}\Gamma_{\rho^0(1450)})}{  ((q^2-m_{\rho^0}^2)^2+q^{2}\Gamma_{\rho^0}^2)((q^2-m_{\rho^0(1450)}^2)^2+q^2\Gamma_{\rho^0(1450)}^2)} \cos{\phi}\\ \nonumber
  &- \frac{2  m_{\rho^0}^2m_{\rho^0(1450)}^2 q^2 (m_{\rho^0}^2+m_{\rho^0(1450)}^2) }{  ((q^2-m_{\rho^0}^2)^2+q^{2}\Gamma_{\rho^0}^2)
  ((q^2-m_{\rho^0(1450)}^2)^2+q^2\Gamma_{\rho^0(1450)}^2)} \cos{\phi}\\  \nonumber
  &+ \frac{2  m^2_{\rho^0} m^2_{\rho^0(1450)}
    (q^3\Gamma_{\rho^0(1450)}+q \Gamma_{\rho^0} m_{\rho^0(1450)}^2)}{((q^2-m_{\rho^0}^2)^2+q^{2}\Gamma_{\rho^0}^2)
    ((q^2-m_{\rho^0(1450)}^2)^2+q^2\Gamma_{\rho^0(1450)}^2)}\sin{\phi}\\ 
  &- \frac{2  m_{\rho^0}^2m_{\rho^0(1450)}^2 (q^3\Gamma_{\rho^0}+q
    \Gamma_{\rho^0(1450)}m_{\rho^0(1450)}^2)}{((q^2-m_{\rho^0}^2)^2+q^{2}\Gamma_{\rho^0}^2)
    ((q^2-m_{\rho^0(1450)}^2)^2+q^2\Gamma_{\rho^0(1450)}^2)}\sin{\phi}\bigg]
\label{eq:CD}
\end{eqnarray}
The other cross terms in \eqref{eq:dsdG} can be obtained by appropriate replacement of the indices in \eqref{eq:CD}.
One can represent  this expression  as a sum of
$(c^{+}_{00}-c^{-}_{00})\cdot\cos{\phi}+(d^{+}_{00}-d^{-}_{00})\cdot\sin{\phi}$,
where c and d are the corresponding terms in formula  \eqref{eq:CD}.
Then expression \eqref{eq:dsdG}  can be rewritten as a sum 
 \begin{equation}
\frac{d\sigma}{d\Gamma} = A + B +  C^{+} \cos{\phi} - C^{-}\cos{\phi} + D^{+}\sin{\phi}-D^{-}\sin{\phi},
\label{eq:dsdGsum1}
\end{equation}
where $A=\bigg|\sum\limits_{j} a_{j}\bigg|^2$, $B= \bigg|\sum\limits_{j} b_{j}\bigg|^2$, $C^{\pm}=\sum\limits_{i,j} c^{\pm}_{ij}$,
and $D^{\pm}=\sum\limits_{i,j} d^{\pm}_{ij}$. 
To calculate detection efficiency, we should simulate separately six 
contributions entering in \eqref{eq:dsdGsum1}.

Similarly, we considered the possible interference of the
$J/\psi\rightarrow \rho\pi$ process with a nonresonant decay into three pions,  $J/\psi\rightarrow \omega\pi$
and  $J/\psi\rightarrow \rho(1700)\pi$. This is discussed in section \ref{sec:fitting}.
The signal MC samples of all contributions are generated for the analysis.

It should be noted that the exact expressions for the amplitudes contain constants
   that are not essential in the MC simulation, but which are important
   in determining the coupling constants.
   The corresponding ratios are as follows \cite{Achasov:1991}:   
   $|g_{J/\psi\gamma}|=\bigg[ \frac{3  m_{J/\psi}^3 \Gamma_{J/\psi} \B(J/\psi\rightarrow e^+ e^-)}{4\pi\alpha}\bigg]^{\frac{1}{2}}$,
   $|g_{J/\psi\rho\pi}|=\bigg[\frac{4\pi\Gamma_{J/\psi}\B(J/\psi\rightarrow\rho\pi)}{W(m_{J/\psi})} \bigg]^{\frac{1}{2}}$,
   where  $W(s)$ is a phase space factor, $\Gamma_{J/\psi}$ is total $J/\psi$ width, $\B(J/\psi\rightarrow e^+ e^-)$ and
   $\B(J/\psi\rightarrow\rho\pi)$ are branching fractions for the mentioned decays.

The KEDR simulation program is based on the GEANT package, version
3.21~\cite{GEANT:Tool}.  The $J/\psi$ decays were simulated with the
BES generator~\cite{BESGEN} based on the JETSET~7.4
code \cite{JETSET} and  tuned in the KEDR experiment \cite{jpsi:2018}.  That allowed us to determine accurately the
number of $J/\psi$    events to obtain the desired branching fractions.
The BHWIDE~\cite{BHWIDEGEN} and MCGPJ generators~\cite{MCGPJ} provided simulation of
$e^+e^-\rightarrow e^+e^-\gamma$ and
$e^+e^-\rightarrow\mu^{+}\mu^{-}\gamma$ events to define the
background from dileptonic processes. To determine hadronic
background, we  simulated the exclusive processes  $ J/\psi \rightarrow
K_{S}^0K^{*}(892)^{0}, K^{*}(892)^{+}K^{-}+c.c.$ with kaons and decay of $J/\psi$  into vector-pseudoscalar
 $J/\psi  \rightarrow \rho\eta, \rho\eta', \phi\eta, \omega\eta, \omega\pi^0$ using generators of the KEDR simulation package.

\section{Experiment and data analysis}\label{sec:data}
The data sample used in this analysis  was taken by the KEDR detector \cite{KEDR:Det} at the VEPP-4M collider
\cite{Anashin:1998sj}.  The process was analysed for a 1.4 pb$^{-1}$
data accumulated at the $J/\psi$ peak consisting of about $5.23 \cdot 10^6$ resonance decays.

\subsection{Event selection}
\begin{figure*}[ht!] 
\centering\includegraphics*[width=0.7\textwidth]{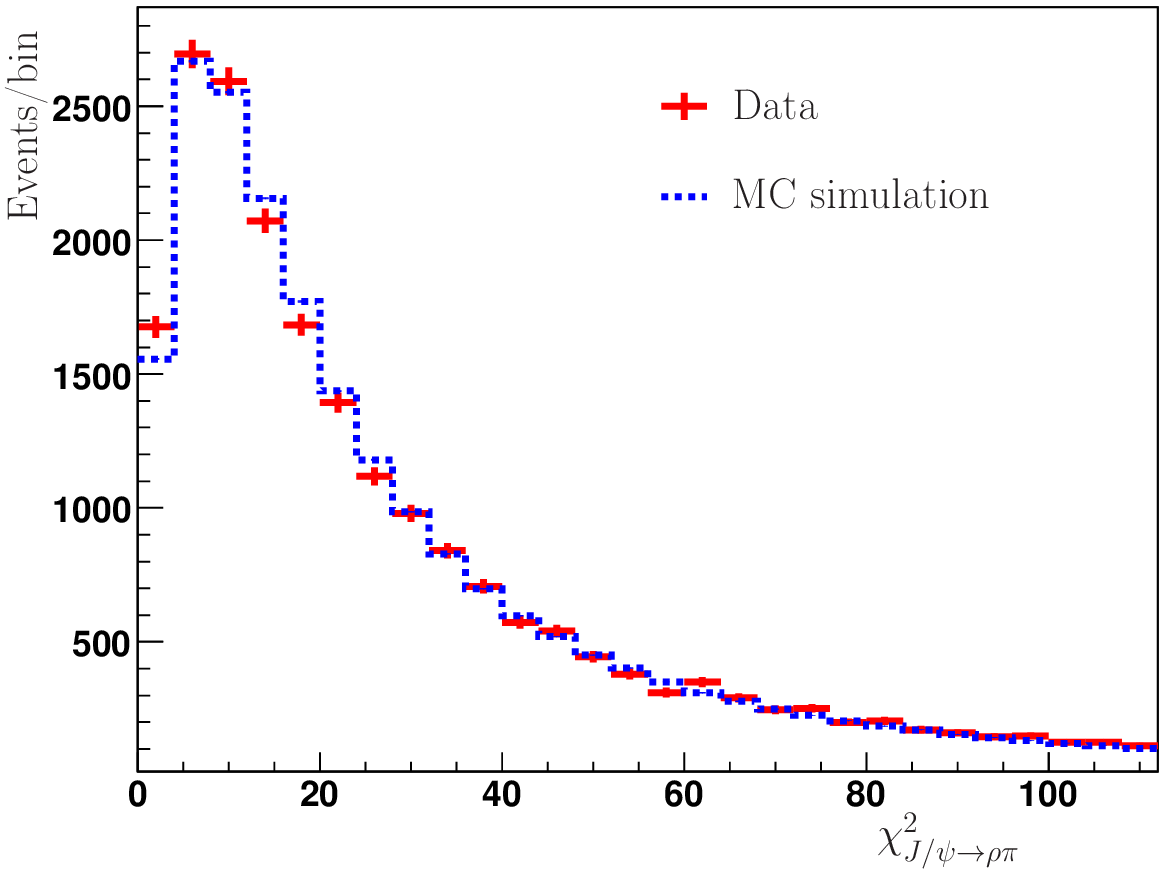}
\caption{ $\chi^2$  distribution of kinematic fit for
  $J/\psi\rightarrow\rho\pi$  selected candidate events. }
\label{f:chi2rhopi}
\end{figure*}

\protect\label{subsec:mhsel}
We select $J/\psi\rightarrow\rho\pi$ events by applying criteria on the track
multiplicity and event topology. 
Two  reconstructed tracks are required to have \mbox{$d<3$}~cm and \mbox{$|z_0|<17$}~cm, 
where $d$ is the track impact parameter relative to the beam axis and
$z_0$ is the coordinate of the closest approach point. 
Only events with at least one track from interaction region
(\mbox{$d<0.75$}~cm,\mbox{$|z_0|<13$}~cm) or two tracks with \mbox{$d<0.75$}~cm   were accepted.
We also required two clusters  in the calorimeter  not  associated  to
tracks ("neutral clusters") with energies exceeding $E_1=50$ MeV or one cluster with an energy greater than $E_2=150$ MeV.  
The selected events are fitted kinematically.
A kinematic fit is applied to reconstruct the candidate events for two
hypotheses: $J/\psi $ decay into $\pi^+\pi^-\pi^0$ and $J/\psi$  decay to
$K^+K^-\pi^0$ in final state.
Neutral pion is reconstructed either from two neutral clusters, 
otherwise from one neutral cluster  ("merged" $\pi^0$) with energy greater than $E_2$.
The kinematic fit adjusts the cluster energy and the track momentum within the
measured uncertainties so as to satisfy energy and momentum
conservation for the given event hypothesis. 
In the case of the merged photons
the momentum conservation condition was not required.
In further selection of events, $\chi^2_{\pi^+\pi^-\pi^0}$  from a
kinematic fit must be less than 90 and also satisfy the condition  $\chi^2_{\pi^{+}\pi^{-}\pi^{0}} < \chi^2_{K^{+}K^{-}\pi^{0}}$.
Figure \ref{f:chi2rhopi} shows the $\chi^2$ distribution of the
kinematic fits for the selected $J/\psi\rightarrow\rho\pi$ events.

\begin{figure*}[h!] 
\centering\includegraphics*[width=0.73\textwidth]{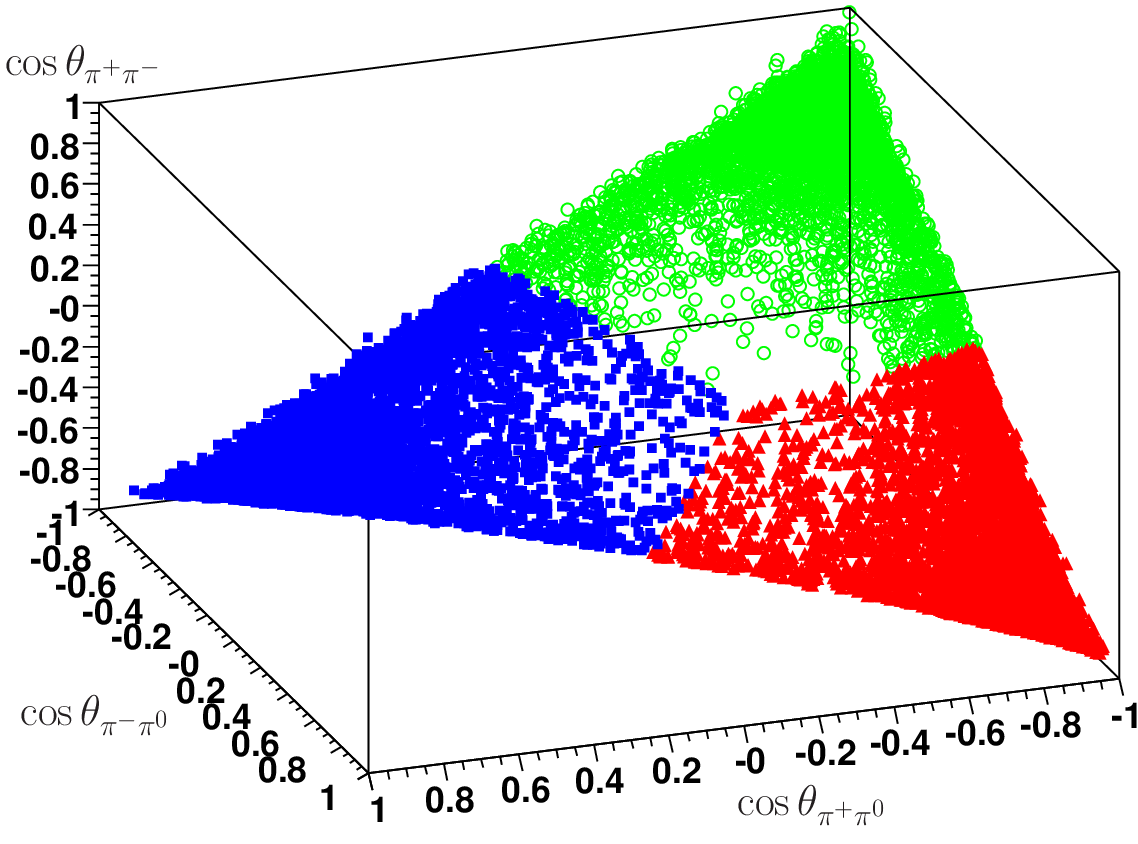}
\caption{The \TCM{ experimental} distribution of events over the cosines 
$\cos{\theta_{\pi^+\pi^0}},\cos{\theta_{\pi^-\pi^0}},\cos{\theta_{\pi^+\pi^-}}$. Triangle,  square and circle  markers correspond to 
the conditions that single out the $J/\psi$ meson decays
to $\rho^+ \pi^-$, $\rho^-\pi^+$ and $\rho^0\pi^0$, respectively.}
\label{f:d3cos}
\end{figure*}

The subsequent stages of the analysis were carried out in accordance
with the ref.~\cite{Todyshev:arx2022}. Three subsets of events are selected 
according to the following conditions:
$\cos{\theta_{\pi^+\pi^0}}\!>\!\cos{\theta_{\pi^+\pi^-}}~\wedge~\cos{\theta_{\pi^+\pi^0}}\!>\!\cos{\theta_{\pi^-\pi^0}}$,
$\cos{\theta_{\pi^-\pi^0}}\!>\!\cos{\theta_{\pi^+\pi^-}}~\wedge~\cos{\theta_{\pi^-\pi^0}}\!>\!\cos{\theta_{\pi^+\pi^0}}$,
 and
$\cos{\theta_{\pi^+\pi^-}}\!>\!\cos{\theta_{\pi^-\pi^0}}~\wedge~\cos{\theta_{\pi^+\pi^-}}\!>\!\cos{\theta_{\pi^+\pi^0}}$. 
Here and below, $\theta_{\pi^+\pi^0}$, $\theta_{\pi^+\pi^-}$ and
$\theta_{\pi^+\pi^-}$ are the angles between the  momentum vectors of
the corresponding $\pi$ mesons. Figure \ref{f:d3cos} 
shows the experimental distribution of these cosines.

For the suppression of the background induced by the processes
$e^+e^-(\gamma)$, $\mu^+\mu^-(\gamma)$ for events with "merged"
$\pi^0$ we used the additional  criteria. 
The ratio of Fox-Wolfram moments \cite{FW:1979} $H_{2}/H_{0}$ was required to be less
than $0.8$. The ratio of the energy deposited in the calorimeter to the measured
momentum of the charged particle  $E/p$  must be less than $0.75$.
The sum $\cos{\theta_{\pi^+\pi^-}}+\cos{\theta_{\pi^+\pi^0}}+\cos{\theta_{\pi^-\pi^0}}$
was required to be less than $-1.075$, this distribution lies in
the range of $-1.5$ to $-1$. 

To reduce events with neutral clusters overlapped from those
associated with track, we also required
$\cos{\theta_{\pi^+\pi^-}}>-0.95$  for events considered as
$J/\psi\rightarrow\rho^{+}\pi^{-}$ and $J/\psi\rightarrow\rho^{-}\pi^{+}$ decays.

\subsection{Analysis procedure}
\protect\label{sec:analysis}
In our analysis, we perform a binned simultaneous fit of the $\rho^0$,
$\rho^+$ and $\rho^-$ invariant mass distributions. The bin sizes are chosen equal to
24 MeV/$c^2$ for the neutral decay mode  and 22 MeV/$c^2$ for the charged decay
modes.
The expected number of events as function of the $\rho$ invariant mass
for given decay mode is parameterized as follows:
\begin{eqnarray}
\label{eq:mdist}
\nonumber
n^{theor}(q) &= \B_{\rho\rightarrow\pi\pi}\B_{\pi^0\rightarrow\gamma\gamma} \bigg[p_1\epsilon_1 H_{\rho\pi} + \epsilon_2  p_2  H_{\rho(1450)\pi} \\\nonumber
                         &+ (\epsilon_3  H^{c+}_{\rho\pi,\rho(1450)\pi} -
                             \epsilon_4
                             H^{c-}_{\rho\pi,\rho(1450)\pi})\cos{\phi}\sqrt{p_1 p_2}\\
                         &+ (\epsilon_5  H^{s+}_{\rho\pi,\rho(1450)\pi}
                            -\epsilon_6
                            H^{s-}_{\rho\pi,\rho(1450)\pi})\sin{\phi}\sqrt{p_1 p_2}\bigg] 
            & + \sum \epsilon_{bkgs}H_{bkgs},
\end{eqnarray}
where   $p_1$ and $p_2$ are parameters related to decays probabilities
and $\phi$ is the interference phase. They are free in the fit. 
$H_{\rho\pi}$,  $H_{\rho(1450)\pi}$, $H^{c\pm}_{\rho\pi,\rho(1450)\pi}$ and
$H^{s\pm}_{\rho\pi,\rho(1450)\pi}$ are the distributions corresponding 
to the terms A, B, C and D in \eqref{eq:dsdGsum1}.  These  distributions
are proportional to the integrals of the functions $A$, $B$,
$C^{\pm}$ and $D^{\pm}$ over the phase space and initially normalized to
$N_{sig}-N_{bkgs}$, where $N_{sig}$  is the number of selected events
of the given $J/\psi\rightarrow\rho\pi$ decay mode and
$N_{bkgs}$ is the expected number of background events.  
The detection efficiencies  $\epsilon_{i}$ and $\epsilon_{bkgs}$ are obtained from the MC simulation.

   For the $J/\psi \rightarrow \rho^{0} \pi^{0}$ decay, the main hadronic  background arises from $J/\psi\rightarrow
K_S^0 K^{*}(892)^0$ $\rightarrow K_S^0 K^{+}\pi^{-}+c.c.$ decays, while for the
$J/\psi\rightarrow \rho^+\pi^-$ and $J/\psi\rightarrow \rho^{-}\pi^{+}$ the dominated part of the hadronic
contamination is due to decays $J/\psi\rightarrow K^{+}K^{*}(892)^{-}\rightarrow K^{+} K_S^0\pi^{-}$ and
$J/\psi\rightarrow K^{-}K^{*}(892)^{+}\rightarrow K^{-} K_S^0\pi^{+}$,
respectively. These contributions, as well as the possible contributions of the QED
processes $ e^+e^-\rightarrow e^+e^-(\gamma)$,
$e^+e^-\rightarrow\mumu(\gamma)$   were simulated and included into
the last term of the fitting function \eqref{eq:mdist}. 
All of them are given in Table~\ref{tab:bcg}. 
The expected number of background events  was estimated 
using the total number $J/\psi$ decays, branching fractions of the background
processes  and their detection efficiencies.

We introduce raw branching fraction $\B^{sig}_{raw}=N_{sig}/(\epsilon_{1} N_{J/\psi} )$,
where $N_{J/\psi}$ is the number of $J/\psi$ events determined with the
equation $N_{J/\psi}=N_{hadr}^{sel}/\epsilon_{J/\psi}$,
$N_{hadr}^{sel}$ is the number  of the selected hadronic $J/\psi$ decays.
The $J/\psi$ detection efficiency  $\epsilon_{J/\psi}$
is derived from the MC simulation.  The product of $p_1$ by $\B^{sig}_{raw}$ allows one to determine the
branching fraction of the decay $J/\psi\rightarrow\rho\pi$ using
selected $J/\psi\rightarrow\rho\pi$ events $\B^{sig}= p_1\cdot\B^{sig}_{raw}$. 
For the branching fraction $\B_{\rho\rightarrow\pi\pi}$ one has $\B_{\rho^0\rightarrow\pi^+\pi^-}=0.98906\pm0.0016,\B_{\rho^{\pm}\rightarrow\pi^{\pm}\pi^0}=0.99955\pm 0.00005 $ and
$\B_{\pi^0\rightarrow\gamma\gamma}=0.98823 \pm 0.00034$ PGD~\cite{PDG:2022}.
The  described approach to fitting distributions was inspired by
the article~\cite{snd:2007}. 
Note that this method differs from the Dalitz plot analysis,
which is used, for example, in \cite{BABAR:2017}.
\begin{table}[h!]
\caption{\label{tab:bcg}  Background contributions to the fit are listed in percent.}
\begin{center}
\begin{tabular}{lccc}                              
  Decay channel                             &  \multicolumn{3}{c}{Modes of the decay $J/\psi\rightarrow\rho\pi$ }   \\
                                             &  $\rho^0\pi^0 $        &  $\rho^+\pi^-$  &   $\rho^-\pi^+ $     \\\hline %
  \multicolumn{4}{c}{ Contribution $e^+e^- \rightarrow e^+e^-(\gamma), \mu^+\mu^-(\gamma)$}   \\\hline 
                     &   $0.2 \pm 0.1$      &    $0.7\pm 0.1$            &     $0.8 \pm 0.1$      \\\hline                     
   \multicolumn{4}{c}{Hadronic contributions }   \\\hline 
  $ K_S^0 K^{*}(892)^0+ c.c.$              &  $0.4  \pm 0.1$  &      $-$              &    $-$            \\\hline   
   $ K^{+}K^{*}(892)^{-}$                 &       $-$          &     $0.5  \pm 0.1$ &    $-$          \\\hline   
   $ K^{-}K^{*}(892)^{+}$                  &       $-$          &           $-$       &   $0.5  \pm 0.1$      \\\hline   
\end{tabular}\\
\end{center}
\end{table}

The observed number of signal events $N_{sig}$, expected number of
background events   $N_{bkgs}$  and related input quantities for all
individual  decay modes  are summarized in  Table~\ref{tab:quant}.
\begin{table}[!h]
\caption{\label{tab:quant}{Summary of the signal and background
    yields, detection efficiencies for each decay mode.
    \TCM{Statistical errors for values $N_{sig},\epsilon_{1\ldots6} $ and total error for $N_{bkgs}$ are indicated.}
     }}
\begin{center}
\begin{tabular}{lccc} 
       Input quantity                    &  \multicolumn{3}{c}{Modes of the decay $J/\psi\rightarrow\rho\pi$ }   \\\hline %
                                         & $J/\psi\rightarrow \rho^0\pi^0$    & $J/\psi\rightarrow \rho^+\pi^-$ & $J/\psi\rightarrow \rho^-\pi^+$  \\\hline %
        $N_{sig}$                           & $5908 \pm 77$    &        $6927\pm83$    &       $6959\pm83$ \\\hline %
        $N_{bkgs}$                          & $34.2\pm 6.6$     & $77.6\pm9.9$   & $91.5\pm10.4$ \\\hline %
       $\epsilon_1, \%$     &  $6.32\pm 0.01$    & $7.08\pm0.01$    & $7.18\pm 0.01$  \\\hline %
       $\epsilon_2, \%$     &  $5.93\pm 0.02$    & $8.39\pm 0.02$   & $8.43\pm 0.02$  \\\hline %
       $\epsilon_3, \%$     &  $6.17\pm 0.02$    & $6.98\pm0.02$    & $7.09\pm0.02$  \\\hline %
       $\epsilon_4, \%$     &  $6.22\pm 0.02$    & $6.97\pm0.02$    & $7.08\pm0.02$  \\\hline %
       $\epsilon_5, \%$     &  $6.22\pm 0.02$    & $7.34\pm0.02$    & $7.43\pm0.02$  \\\hline %
       $\epsilon_6, \%$     &  $6.21\pm 0.02$    & $7.11\pm0.02$    & $7.21\pm0.02$  \\\hline %
\end{tabular}\\
\end{center}
\end{table}

The numbers of $J/\psi\rightarrow\rho\pi$ events observed at each
decay modes $j$ and each invariant mass interval $k$
were fitted simultaneously as a function of invariant mass using a minimizing function\\
{\large
\begin{equation}
\chi^2 =\sum\limits_{j} \sum\limits_{k} \frac{(n_{jk}^{\text{exp}}-n_{jk}^{\text{theor}})^2}{n_{jk}^{\text{exp}}+\sigma^2_{n_{jk}^{\text{theor}}}}
\label{chi2fit}
\end{equation}
}
where $n_{jk}^{\text{exp}}$ and $n_{jk}^{\text{theor}}$ are 
experimentally measured and theoretically calculated numbers of $J/\psi\rightarrow\rho\pi$ events, respectively.
\TCM{$\sigma_{n_{jk}^{\text{theor}}}$ is error of the calculated $n_{jk}^{\text{theor}}$.}

\begin{figure*}[ht!] 
\begin{center}
\centering\includegraphics*[width=0.95\textwidth]{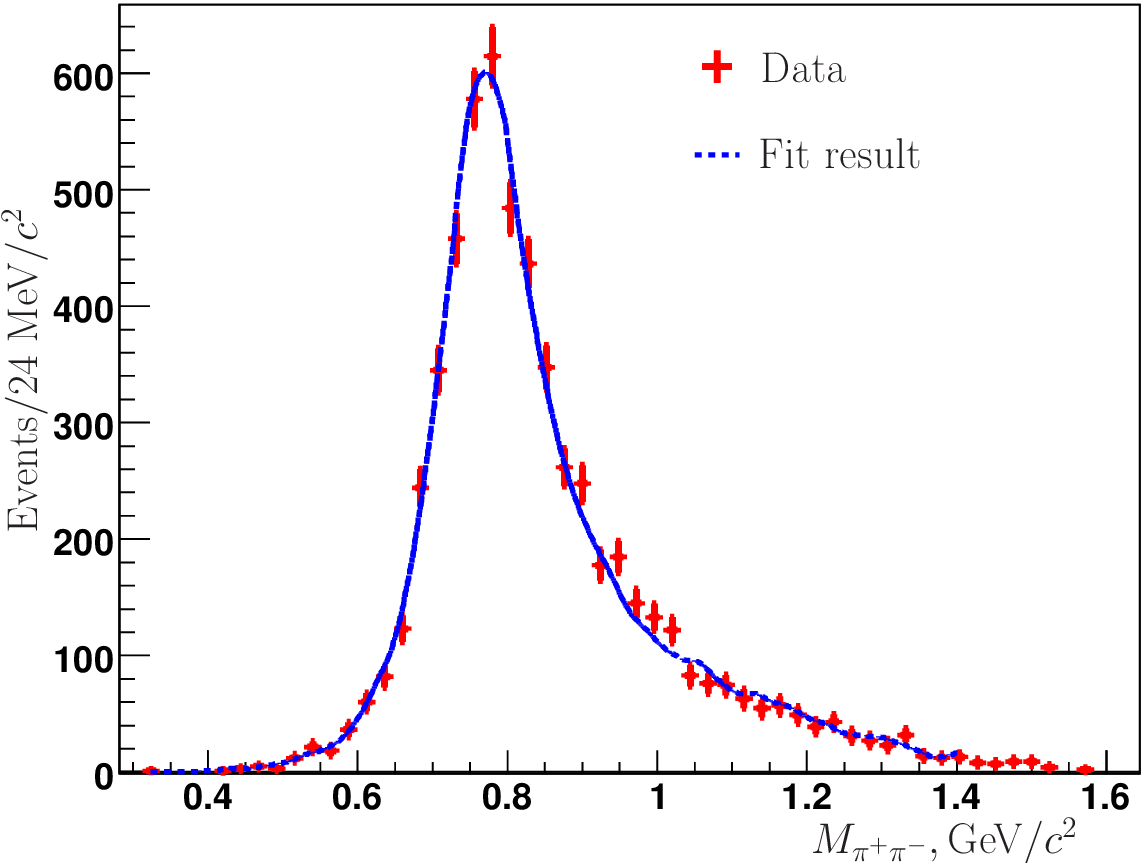}\\[1pc]
\centering\includegraphics*[width=0.49\textwidth]{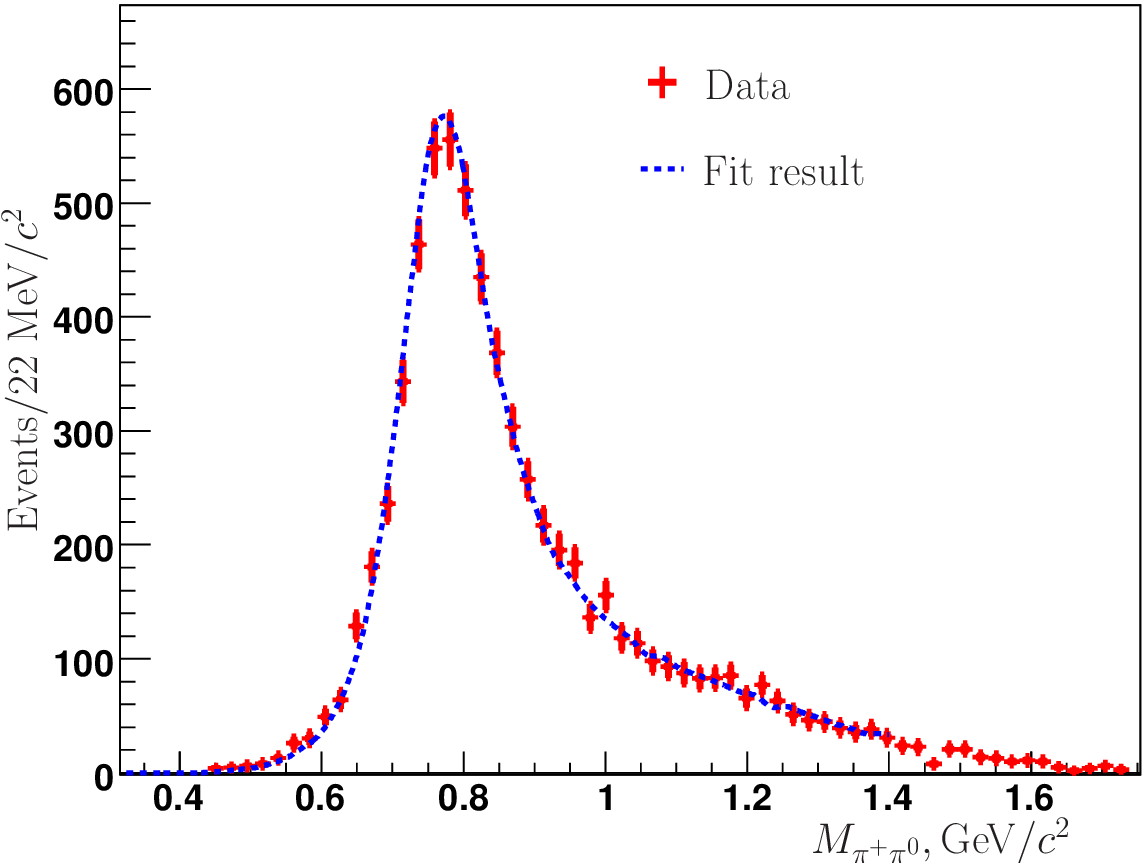}
\centering\includegraphics*[width=0.49\textwidth]{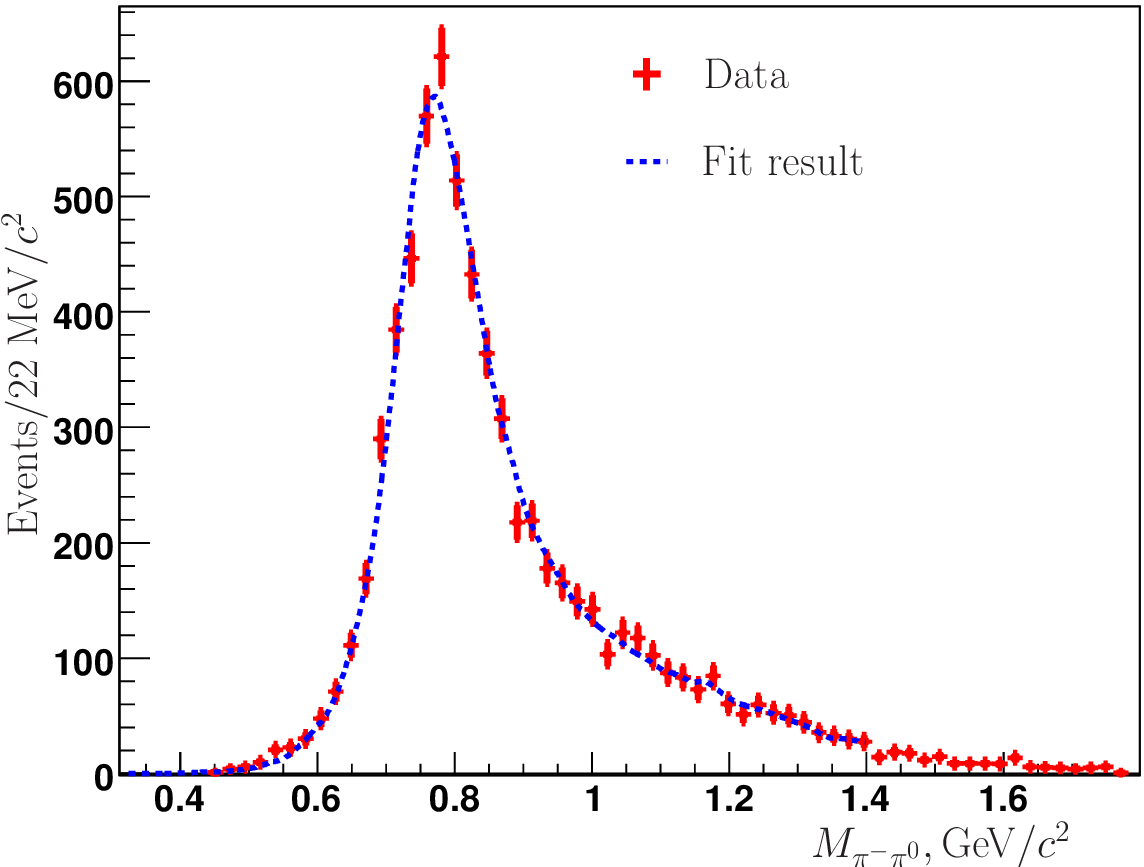}
\caption{The invariant masses distributions of $\pi^+\pi^-$,  $\pi^+\pi^0$ and  $\pi^-\pi^0$. The dashed curve shows  the result of the simultaneous fit. The experimental data set is presented in tables \ref{tab:app1} and \ref{tab:app2} in Appendix.}
\label{f:mdist}
\end{center}
\end{figure*}

Figure \ref{f:mdist} shows the result of the  fit of the $\rho$
meson's invariant mass distributions over all decay modes $J/\psi\rightarrow\rho^0\pi^0$, $J/\psi\rightarrow\rho^+\pi^-$ and $J/\psi\rightarrow\rho^-\pi^+$.

The fitting was carried out in the range of invariant masses up to 1.4 GeV/c$^2$.
The first three free parameters  determine the branching fraction  
$\B(J/\psi\rightarrow\rho\pi)$ based on subsets of events 
$J/\psi\rightarrow\rho^{+}\pi^{-}$, $J/\psi\rightarrow\rho^{-}\pi^{+}$ and $J/\psi\rightarrow\rho^{0}\pi^{0}$ modes.
We will denote these parameters as $\B^+$, $\B^-$ and $\B^{0}$, respectively.
The parameters defining \TCM{the products $\B_{\rho^{0}(1450)\pi^{0}}\cdot\B(\rho(1450)\rightarrow\pi^+\pi^-)$,
$\B_{\rho^{+}(1450)\pi^{-}}\cdot\B(\rho(1450)\rightarrow\pi^+\pi^0),\B_{\rho^{-}(1450)\pi^{+}}\cdot\B(\rho(1450)\rightarrow\pi^-\pi^0)$} and the phase of the interference are
also free, but we considered them as just auxiliary quantities.
We took into account the possible shift of the invariant mass between
experiment and simulation by introducing an additional free parameter
$\delta M$. 
The function $n^{theor}(q)$ is defined for all possible values of $q$,
since a cubic spline approximation is constructed over the entire
range of invariant masses.
The   branching fractions of the process $J/\psi\rightarrow\rho\pi$
obtained from the fit are presented in  Table~\ref{tab:fit}. 
\begin{table}[h!]
\begin{center}
\caption {\label{tab:fit} {\normalsize The main results of simultaneous fit
    (only statistical errors are presented).}}
\begin{tabular}[l]{|c|c|c|c|c|c|c|} \hline
   $\B^{+},\%$     &  $\B^{-},\%$    & $\B^{0},\%$        & $\delta   M$, MeV/c$^2$&$\phi$, rad  & $\chi^2/ndf$  & $P(\chi^2)$ \\\hline
   $2.028\pm 0.029$  &  $2.017\pm0.027$ & $2.053\pm0.030$ & $5.2 \pm   0.8$  &$-1.89\pm 0.13$ & $122.2/121$   & 0.45  \\\hline
\end{tabular}
\end{center}

\end{table}

Based on the fit results obtained, we determined the average
value  $\B(J/\psi\rightarrow\rho\pi)=(2.031 \pm 0.017)\cdot 10^{-2}$.
This result is given without corrections, which are discussed in 
sections \ref{sec:fitting} and \ref{err:deterr}.
\TCM{ Obtained similarly from the fitting results, the product
  $\B(J/\psi\rightarrow\rho(1450)\pi)\cdot\B(\rho(1450)\!\rightarrow\!\pi\pi)$ is equal to
  $(1.88\pm0.22)\cdot 10^{-4}$.
At the same time, the contribution of the destructive interference of processes $J/\psi\rightarrow\rho\pi$ and
$J/\psi\rightarrow\rho(1450)\pi$  to the observed cross section is approximately $-11.9\%$.
The issue of determining the quantity
$\B(J/\psi\rightarrow\rho(1450)\pi)\cdot\B(\rho(1450)\rightarrow\pi\pi)$ is described in more detail in Section \ref{sec:rho1450}. 
}

\section{Discussion of systematic uncertainties in  $\B(J/\psi\rightarrow\rho\pi)$}
\subsection{Systematic uncertainty of the fitting model}
\protect\label{sec:fitting}
The inaccuracy of $\rho(1450)$ resonance parameters introduces
uncertainty to the branching fraction obtained from the fit.
This uncertainty is evaluated by the variation of the
 $\Gamma_{\rho(1450)}$ and $M_{\rho(1450)}$ in the ranging of their
errors 60 MeV and 25 MeV, respectively, taken from PDG~\cite{PDG:2022}.
The resulting changes of the $J/\psi\rightarrow\rho\pi$ 
branching fraction were $1.0\%$ and $0.2\%$.
\TCM{The uncertainty related to the parameters of the
  $\rho(770)$  meson is due to an inaccuracy of 0.8 MeV in determining 
its total width \cite{PDG:2022} and is estimated at $0.5\%$ 
in a similar way.}
 

The possible contribution of the process $J/\psi\rightarrow\gamma f_2$  
was simulated and included into the fit. The change of the measured
branching fraction is $-0.2\%$. We apply this correction to the our
result and include an additional  $0.1\%$ error into the systematic
uncertainty.

In equation \eqref{eq:dsdG}, the contributions related to
$J/\psi\rightarrow\omega\pi$, $J/\psi\rightarrow\rho(1700)\pi$ processes  and
nonresonant three-pion decay are omitted. 
The systematic uncertainties associated with this approximation 
were estimated by adding these terms one by one to equation
\eqref{eq:dsdG} similarly to  $J/\psi\rightarrow\rho(1450)\pi$
contribution as described in Section \ref{sec:frame}. 
In each case two additional free parameters were introduced, the
amplitude of the process and the interference phase.  

The systematic uncertainties obtained are presented in Table \ref{tab:ferr}.
We also took into account the MC statistical uncertainty and the systematic errors related to uncertainties in the
$\B_{\rho^0\rightarrow\pi^+\pi^-}$, $\B_{\rho^{\pm}\rightarrow\pi^{\pm}\pi^0}$
and $\B_{\pi^0\rightarrow\gamma\gamma}$ parameters entering in \eqref{eq:mdist}.
\begin{table}[h!]
\caption{\label{tab:ferr} The relative systematic uncertainties in
  $\B(J/\psi\rightarrow\rho\pi)$ due to approximation  in the invariant mass distribution.}
\begin{center}
\begin{tabular}{lcc}
    Source                                     & Uncertainty, $\%$    \\\hline                                         
Uncertainty $\Gamma_{\rho(1450)}$              &  $1.0$        \\ %
Uncertainty $M_{\rho(1450)}$                   &  $0.2$    \\ %
\TCM{Uncertainty $\Gamma_{\rho(770)}$ }              &  \TCM{~$0.5$  }  \\ %
Contribution of $\gamma f_2$                   &  $0.1$      \\                              %
Contribution $\rho(1700)\pi$                    & $0.7$      \\ %
Contribution   $e^+e^- \rightarrow\pi^{+}\pi^{-}\pi^0$ & $0.6$     \\ 
Contribution $\omega\pi$                     & $0.2$  \\ %
MC statistics                                  & $0.2$ \\ %

Uncertainties $\B_{\rho^0\rightarrow\pi^+\pi^-}$,
$\B_{\rho^{\pm}\rightarrow\pi^{\pm}\pi^0}$, $\B_{\pi^0\rightarrow\gamma\gamma}$  & $0.1$    \\\hline %
Sum in quadrature                              & $1.5$  \\\hline %
\end{tabular}
\end{center}
\end{table}

\subsection{Systematic uncertainty of  the fitting procedure}
\protect\label{sec:loc}    
Since we perform a simultaneous fitting of the $\rho$ meson's
invariant mass distributions the results obtained are sensitive to the
method of delimiting decay modes. 
To estimate this uncertainty we considered the alternative method of
modes  separation in accordance with the conditions $\cos{\theta_{\pi^+\pi^-}} \!<\! P \wedge
\cos{\theta_{\pi^+\pi^0}} \!>\!\cos{\theta_{\pi^-\pi^0}}$,
$\cos{\theta_{\pi^+\pi^-}} \!<\! P \wedge
\cos{\theta_{\pi^-\pi^0}}\!>\!\cos{\theta_{\pi^+\pi^0}}$ and $\cos{\theta_{\pi^+\pi^-}}> P $ to select
$J/\psi\rightarrow\rho^+\pi^-$, $J/\psi\rightarrow\rho^-\pi^+$ and
$J/\psi\rightarrow\rho^0\pi^0$ events, respectively. The $P$ parameter
varied from $-0.45$ to $-0.55$. 
We also modified the method described earlier by replacing "$\cos{\theta_{\pi^+\pi^-}}$" with  "$\cos{\theta_{\pi^+\pi^-}}+\delta$" in all conditions listed in
Section \ref{subsec:mhsel}. In our fit, we used the value of $\delta=-0.3$
according to ref.~\cite{Todyshev:arx2022}. This value corresponds to the
minimum intersection of the sets of events of the considered decay
modes. 
The maximum change of the measured  $J/\psi\rightarrow\rho\pi$
branching fraction for these two methods was  $1.1\%$.

We have also varied the invariant mass ranges.
When varying the upper invariant mass limit of the fit from $1.3$ GeV/c$^2$
to $1.8$ GeV/c$^2$, difference in the obtained $J/\psi\rightarrow\rho\pi$
branching fraction were less than  $0.4\%$.

Systematic uncertainties  described in this section are given in Table \ref{tab:perr}.
\begin{table}[h!]
\caption{\label{tab:perr} The  systematic uncertainties for $\B(J/\psi\rightarrow\rho\pi)$ associated with fitting procedure.}
\begin{center}
\begin{tabular}{lcc}
    Source  & Uncertainty, $\%$    \\\hline %
Variation of the modes separation                                & $1.1$    \\ %
Variation of the fit energy range                                & $0.4$       \\\hline  %
Sum in quadrature                                                & $1.2$  \\\hline %
\end{tabular}
\end{center}
\end{table}

\subsection{Systematic uncertainty in the number of  $J/\psi$ events}
\protect\label{sec:mchadrerr}

\begin{figure*}[ht!] 
\centering\includegraphics*[width=0.6\textwidth]{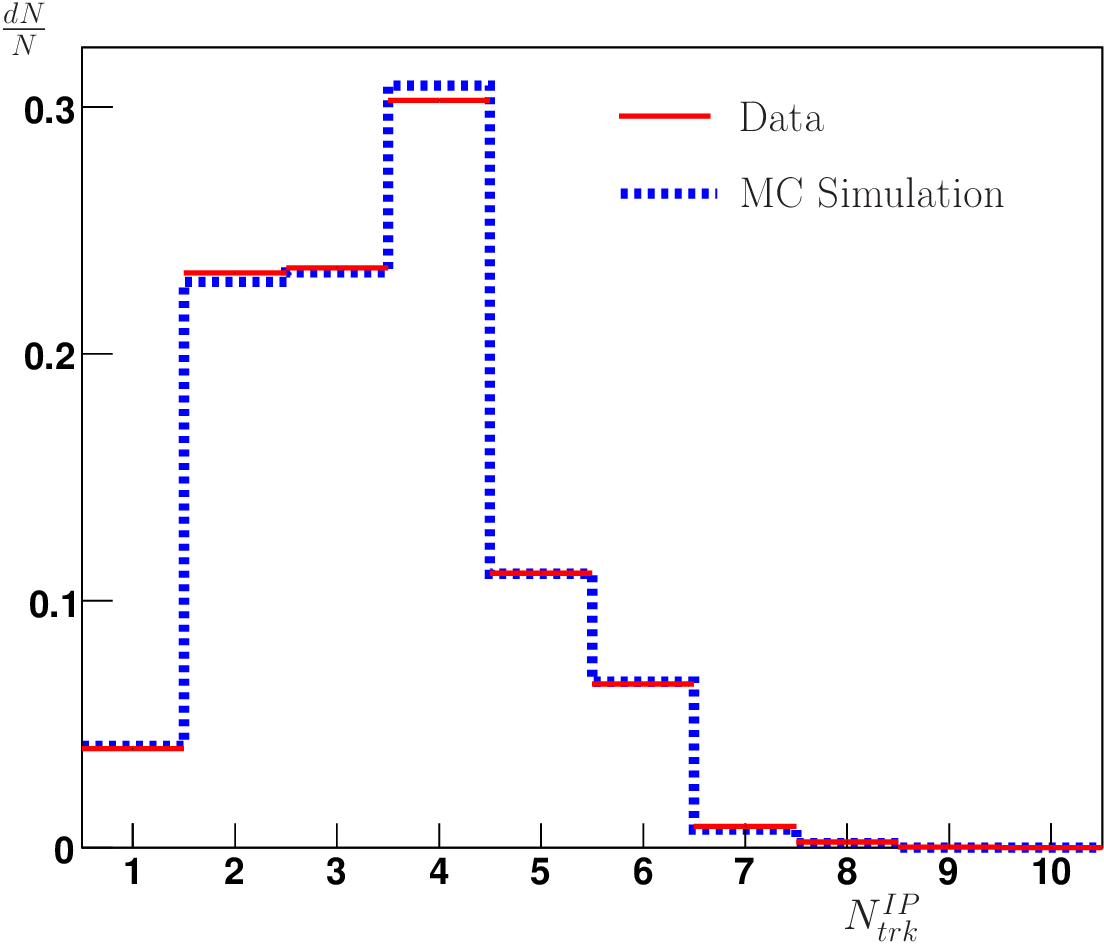}\\
\caption{Distribution of the number of tracks from the interaction point at the $J/\psi$ peak. Distribution is normalized to unity.
}
\label{f:ntjpsi}
\end{figure*}

The details of the Monte-Carlo $J/\psi$ decay simulation and the
procedure  a reliable systematic uncertainty estimation are
described in ref.~\cite{jpsi:2018}. 
Figure \ref{f:ntjpsi} shows comparison between $J/\psi\rightarrow hadrons$
data and the MC simulation for the distribution of the number 
of tracks from the interaction point.
According to this work the error associated with the multihadron event $J/\psi$  generator is about $0.7\%$.

Taking into account the change in the condition of the detector compared
to 2005, additional tuning of the $J/\psi$ decay simulation was carried
out. As a result, the detection efficiency of the multihadron events changed by $0.4\%$.

In addition, we varied criteria for the hadron selection to evaluate
the effect of other possible sources of a systematic uncertainty.
The sum in quadrature of all errors obtained by the
variation of the selection criteria is about $0.8\%$.

Summing up in quadratures these three values, we obtain that 
the conservative error in determining the branching fraction due to
the uncertainty  in the number of $J/\psi$ decays is $1.1\%$.
\subsection{Physical background}
The main background contributions are summarized in Table \ref{tab:bcg}.
The contribution of other background processes such as  $\rho \eta$,
$\rho \eta'$, $\phi \eta$, $\omega\eta$, $\omega\pi^0$, which
are not accounted into the fit was estimated  to be below $0.1\%$ by the 
Monte-Carlo simulation.
Thus, we get the total uncertainty due to background processes estimated of about $0.2\%$.

\subsection{Detector-related uncertainties}
\protect\label{err:deterr}
The track reconstruction efficiency was studied by
$J/\psi\rightarrow\rho^{+}\pi^{-}$ and
$J/\psi\rightarrow\rho^{-}\pi^{+}$ events with reconstructed $\rho$ meson.
About $7.3 \cdot 10^3$ events were selected. In $1.11 \pm 0.12\%$
of cases, the track corresponding to the charged $\pi$ meson was missed. According to the simulation, the
fraction of such events was $0.53\pm0.01\%$, that corresponds to  the
difference $0.58\%$  of the track reconstruction efficiencies. The change of this value does not exceed
$0.22\%$ with a significant tightening of the conditions on the $\rho$ meson invariant mass.
That allow us to introduce  correction $+1.16\pm 0.24\pm 0.44\%$ to the measured branching fraction.
Considering $J/\psi\rightarrow\rho^{0}\pi^{0}$ process events with
reconstructed $\rho^0$ meson, we determined the correction of  $+1.02\pm 0.12 \pm 0.18\%$  to the branching fraction due to missing $\pi^0$.

To estimate the systematic uncertainty related to the momentum and
angular resolution, two methods were used to achieve agreement
between the data and the MC simulation: 
we scale either the assumed systematic errors in x(t) or the drift
chamber spatial resolution. The difference $0.5\%$ between  results
obtained is taken as the systematic uncertainty estimate.

The trigger and event selection efficiencies are sensitive to the nuclear
interaction of pions in the detector material.
We estimated the uncertainty of $0.4\%$ comparing the detection efficiencies for the
$J/\psi\rightarrow\rho\pi$ decay obtained with the
packages GHEISHA~\cite{Fesefeldt:1985yw} and FLUKA~\cite{Fasso:2005zz} 
implemented in GEANT~3.21~\cite{GEANT:Tool}. 

The total correction of the measured branching fraction related to
detector response is $+2.2\%$  with the uncertainty of about $0.8\%$.
The corresponding contributions  are listed in Table~\ref{tab:detresp}.

\begin{table}[h!]
\caption{\label{tab:detresp} {Detector-related uncertainties in $\B(J/\psi\rightarrow\rho\pi)$.}}
\begin{center}
\begin{tabular}{lc}      
   Source            &   Uncertainty, $\%$  \\\hline  
 Track reconstruction                       &        0.5  \\
 $\pi^0$ reconstruction                     &        0.2  \\         
 Tracking $p/\theta$ resolution             &        0.5  \\         
 Nuclear interaction                        &        0.4  \\ \hline        
 Sum in quadrature                          &        0.8  \\                                   
\end{tabular}
\end{center}
\end{table}

The effect of other possible sources of the detector-related uncertainty 
was evaluated by varying the event selection criteria as presented in
Table~\ref{tab:criteriavar}.
\TCM{The observed variation in the number of selected events was significant, with a change in the condition for the $\chi^2_{\pi^{+}\pi^{-}\pi^{0}}$ criteria, it was about $10\%$, and in the absence of the condition $\chi^2_{\pi^{+}\pi^{-}\pi^{0}} <\chi^2_{K^{+}K^{-}\pi^{ 0} }$  reached $40\%$.}
The variations of result can originate from
the already considered sources and statistical fluctuations, 
nevertheless we included them in the total uncertainty to obtain conservative error estimates. 

\begin{table}[h!]
\caption{\label{tab:criteriavar} {$\B(J/\psi\rightarrow\rho\pi)$ uncertainties due to variation of the selection criteria.}}
\begin{center}
\begin{tabular}{llc}      
 Condition$/$Variable            &   Range  variation & Variation   \\         
                                 &   & $\B(J/\psi\rightarrow\rho\pi)$ in $\%$  \\\hline  
\TCM{ $\chi^2_{\pi^{+}\pi^{-}\pi^{0}}$   }   & \TCM{   $<\chi^2_{K^{+}K^{-}\pi^{0}}$ or no cut   } & \TCM{ 1.2}  \\
 $\chi_{\pi^{+}\pi^{-}\pi^{0}}^2$         &     $< ~~70 \div 110$     & 0.3  \\
 $E_{1}$           &      $< ~~40 \div 80$~MeV       & 0.2  \\         
 $E_{2}$           &      $< ~~140 \div 180$~MeV     & 0.1  \\           
 $\cos{\theta_{\pi^+\pi^-}}$(for $\rho^{\pm}{\pi^{\mp}}$ modes)  & $ > -0.995 \div -0.900 $       & 1.1    \\         
$\cos{\theta_{\pi^+\pi^-}}+\cos{\theta_{\pi^+\pi^0}}+\cos{\theta_{\pi^-\pi^0}}$   &   $<-1.15\div 1.05$ & 0.7   \\ 
 $E/p$                   &     $< ~~~0.7  \div 0.8$  & 0.3   \\ 
 $H_2/H_0$               &     $< ~~0.75 \div 0.85$  & 0.1    \\ \hline 
\multicolumn{2}{c}{Sum in quadrature}               & 1.8   \\                             
\end{tabular}
\end{center}
\end{table}

\subsection{Summary of systematic uncertainties}
\protect\label{sec:errsummary}
The main sources of the systematic uncertainty on the measured
branching fraction  are listed in Table~\ref{tab:rerr}.
\renewcommand{\arraystretch}{1.2}
\begin{table}[h!]
\caption{\label{tab:rerr} Dominant systematic uncertainties in the $\B(J/\psi\rightarrow\rho\pi)$.}
\begin{center}
\begin{tabular}{lc}
  Source   &  Uncertainty, $\%$ \\\hline %
  Fitting model             & $1.5$     \\ %
  Fitting procedure         & $1.2$     \\ %
  Number of $J/\psi$ decays & $1.1$    \\ %
  Detector response         & $0.8$        \\%
  Background                & $0.2$     \\ %
  Selected criteria         & $1.8$      \\\hline %
Sum in quadrature           & $3.0$       \\\hline               
\end{tabular}
\end{center}
\end{table}

\section{Determination of the branching fraction of  $J/\psi\rightarrow\rho(1450)\pi$ and $J/\psi\rightarrow\pi^+\pi^-\pi^0$}
\protect\label{sec:rho1450}
\TCM{
The fit performed in the region of invariant masses up to 1.4~GeV/$c^2$, as described in Section \ref{sec:analysis}, 
makes it possible to determine the value of $\B(J/\psi\rightarrow\rho(1450)\pi)\cdot\B(\rho(1450)\rightarrow\pi\pi)$.
However, for a more reliable determination of this value, it is
necessary to expand the fitting region to 1.8~GeV/c$^2$. In this case,
the corresponding contributions from the $J/\psi\rightarrow\rho(1700)\pi$ and $J/\psi\rightarrow\gamma f_2$
processes must be included in the fit.
These changes result in the following value $\B(J/\psi\rightarrow\rho(1450)\pi)\cdot\B(\rho(1450)\rightarrow\pi\pi)=(2.17\pm 0.20)\cdot 10^{-4}$.
The dominant systematic errors in determining the value of 
$\B(J/\psi\rightarrow\rho(1450)\pi)\cdot\B(\rho(1450)\rightarrow\pi\pi)$ are given in Table~\ref{tab:rho1450}. 
\begin{table}[h!]
\caption{\label{tab:rho1450} The relative systematic uncertainties in  $\B(J/\psi\rightarrow\rho(1450)\pi)\cdot\B(\rho(1450)\rightarrow\pi\pi)$.}
\begin{center}
\begin{tabular}{lcc}
    Source                                        & Uncertainty, $\%$    \\\hline                                         
Uncertainty    $\Gamma_{\rho(1450)}$              &  $~9$        \\ %
Uncertainty    $M_{\rho(1450)}$                   &  $~7$    \\ %
Contribution   $\rho(1700)\pi$                    &  $35$      \\ %
Contribution   $e^+e^- \rightarrow\pi^{+}\pi^{-}\pi^0$ & $33$     \\ 
Variation of the modes separation                  &  $15$  \\\hline        
Sum in quadrature                                  &  $52$  \\\hline %
\end{tabular}
\end{center}
\end{table}
}

\TCM{
Based on the observed number of three pion decays
$J/\psi\rightarrow\pi^+\pi^-\pi^0$  over the entire range of
invariant masses of pion pairs and the calculated values of the contributions $J/\psi\rightarrow\rho\pi$,
$J/\psi\rightarrow\rho(1450)\pi$ and the possible contributions of the other processes,
we can calculate the weighted efficiency and the quantity
$\B(J/\psi\rightarrow\pi^+\pi^-\pi^0)=(1.841\pm 0.013) \cdot 10^{-2}$.
In a conservative approach, the systematic uncertainties of a given
value included in category "Fitting model" do not exceed similar
errors in Table \ref{tab:ferr}. 
Estimates of all other systematic uncertainties are the same as for
the $J/\psi\rightarrow\rho\pi$ process, except for the uncertainties
indicated in the \ref{sec:loc} section, which, for obvious reasons,
are absent. Thus, the quadratic sum of systematic uncertainties is $2.7\%$.
}



\section{Summary}
\protect\label{sec:summary}
The measurement of the  $J/\psi\rightarrow\rho\pi$  branching fraction
is performed using  the  data sample of 1.4~pb$^{-1}$ collected at the
$J/\psi$ resonance peak with the KEDR detector. \TCM{The results are
  $\B(J/\psi\rightarrow\rho\pi)  =  (2.072 \pm 0.017\pm 0.062) \cdot 10^{-2}$,
  $\B(J/\psi\rightarrow\rho(1450)\pi) \cdot\B(\rho(1450)\rightarrow\pi\pi)=(2.2\pm 0.2 \pm 1.1) \cdot 10^{-4}$,
  and $\B(J/\psi\rightarrow\pi^+\pi^-\pi^0)= (1.878 \pm 0.013\pm 0.051) \cdot 10^{-2}$
  where the first uncertainty is statistical and the second one is systematic.}
Our results include the correction factor $1.020$ due to the effects
described in the sections \ref{sec:fitting} and \ref{err:deterr}.
These  are the most precise measurements  of $\B(J/\psi\rightarrow\rho\pi)$ and  $\B(J/\psi\rightarrow\pi^+\pi^-\pi^0)$ to date. 

We observe substantial discrepancy with respect to the previous
experiments \cite{BES:2012,BaBar:2007,BaBar:2021} for the $\B(J/\psi\rightarrow\pi^+\pi^-\pi^0)$ value.
We believe that the discrepancy with \cite{BES:2012} are due to the fact
in the present work we employ a more accurate parametrization of the $\pi-\pi$ invariant mass disribution including
interference with $\rho(1450)\pi$ decay.
The result \cite{BaBar:2007} should be corrected taking into account changes in the measurement results of the branching fraction
$\B(\psi(2S) \rightarrow J/\psi \pi^+ \pi^-)$ which has changed by about ten percent.
In addition, the method of Dalitz plot analysis of the ISR events was used in works \cite{BaBar:2007,BaBar:2021}
and the result of our work is based on the data obtained when collecting statistics at the resonance peak.

It should be noted that in Refs.\cite{MRK1:1976,CNTR:1976,DASP:1978,PLUT:1978,MRK2:1983,MRK3:1988,BES:1996,BES:2004,BABAR:2004}, decay $\B(J/\psi\rightarrow\rho\pi)$ was assumed to dominate in the three-pion process, and branching $\B(J/\psi\rightarrow\pi^+\pi^-\pi^0)$ was actually measured.
With this in mind, if we  formally average the results of all experiment excluding BaBar results,
we get  $\B(J/\psi\rightarrow\pi^+\pi^-\pi^0)=(1.88 \pm 0.13)\cdot 10^{-2}$ (error includes scale factor 2.8)
and this is consistent with the result of our work.

We also note the result obtained for the $\B(J/\psi\rightarrow\rho\pi)$ branching fraction has become 
closer to the theoretical calculation given in \cite{Flores-Baez:2016}, 
which, taking into account the change in experimental data, gives the value $\B(J/\psi\rightarrow\rho\pi)=1.74 \cdot 10^{-2}$.

\section*{Acknowledgments}
We greatly appreciate the efforts of the staff of VEPP-4M to provide
good operation of the complex during long term experiments.
The authors are grateful to V.P.~Druzhinin and A.I.~Milstein for useful discussions.
The Siberian Supercomputer Center and Novosibirsk State University Supercomputer
Center are gratefully acknowledged for providing supercomputer facilities.

\newpage
\section{Appendix}
\renewcommand{\arraystretch}{0.75}
\begin{table}[h!]
\footnotesize
\caption{\label{tab:app1} The numbers of the selected $J/\psi\to\rho^0\pi^0$ events and estimated number of the background events 
   for the $\rho^0$ invariant mass distribution.} 
\begin{center}
  \begin{tabular}{lcc}
    $M_{\pi^+\pi^-}$, GeV & $n^{exp}$ & $n_{bkgs}$  \\ \hline
0.324 & 1   & 0.007 \\ \hline
0.420 & 1   & 0.061 \\ \hline
0.444 & 3   & 0.136 \\ \hline
0.468 & 5   & 0.154 \\ \hline
0.492 & 3   & 0.135 \\ \hline
0.516 & 12  & 0.269 \\ \hline
0.540 & 22  & 0.448 \\ \hline
0.564 & 19  & 0.719 \\ \hline
0.588 & 37  & 0.587 \\ \hline
0.612 & 61  & 0.908 \\ \hline
0.636 & 83  & 1.244 \\ \hline
0.660 & 125 & 1.993 \\ \hline
0.684 & 247 & 2.961 \\ \hline
0.708 & 349 & 4.131 \\ \hline
0.732 & 464 & 6.105 \\ \hline
0.756 & 583 & 5.061 \\ \hline
0.78  & 616 & 1.341 \\ \hline
0.804 & 485 & 0.711 \\ \hline
0.828 & 437 & 0.546 \\ \hline
0.852 & 348 & 0.531 \\ \hline
0.876 & 262 & 0.356 \\ \hline
0.900 & 248 & 0.273 \\ \hline
0.924 & 178 & 0.392 \\ \hline
0.948 & 185 & 0.281 \\ \hline
0.972 & 145 & 0.156 \\ \hline
0.996 & 133 & 0.159 \\ \hline
1.020 & 122 & 0.132 \\ \hline
1.044 & 83 & 0.067 \\ \hline
1.068 & 76 & 0.085 \\ \hline
1.092 & 75 & 0.255 \\ \hline
1.116 & 63 & 0.030 \\ \hline
1.140 & 55 & 0.292 \\ \hline
1.164 & 58 & 0.850 \\ \hline
1.188 & 49 & 0.011 \\ \hline
1.212 & 39 & 0.240 \\ \hline
1.236 & 43 & 0.004 \\ \hline
1.260 & 31 & 0.000 \\ \hline
1.284 & 27 & 0.009 \\ \hline
1.308 & 23 & 0.000 \\ \hline
1.332 & 34 & 2.275 \\ \hline
1.356 & 14 & 0.212 \\ \hline
1.380 & 12 & 0.005 \\ \hline
1.404 & 13 & 0.009 \\ \hline
1.428 & 8  & 0.014 \\ \hline
1.452 & 7  & 0.014 \\ \hline
1.476 & 9  & 0.006 \\ \hline
1.500 & 9  & 0.017 \\ \hline
1.524 & 4  & 0.011 \\ \hline
1.572 & 2  & 0.002 \\ \hline
  \end{tabular}
\end{center}
\end{table}

\begin{table}[h!]
\footnotesize
\caption{\label{tab:app2}  The numbers of the selected $J/\psi\to\rho^{-}\pi^+$ and  $J/\psi\to\rho^{+}\pi^-$ events and estimated number of the background  events for the corresponding $\rho$ invariant mass distributions.}
\begin{center}
  \begin{tabular}{lcccc}
                   &   \multicolumn{2}{c}{$J/\psi\rightarrow\rho^{-}\pi^+$}&  \multicolumn{2}{c}{$J/\psi\rightarrow\rho^{+}\pi^-$}  \\\hline              
    $M_{\pi^{-}\pi^0}/M_{\pi^{+}\pi^0}$, GeV &  $n^{exp}$ &  $n_{bkgs}$ & $n^{exp}$ & $n_{bkgs}$\\\hline
0.451 & 1 & 0 & 4 & 0.089 \\ \hline
0.473 & 4 & 0.023 & 5 & 0.393 \\ \hline
0.495 & 7 & 1.130 & 6 & 0.002 \\ \hline
0.517 & 11 & 1.300 & 8 & 0.346 \\ \hline
0.539 & 21 & 0.461 & 13 & 0.167 \\ \hline
0.561 & 23 & 0.358 & 29 & 2.894 \\ \hline
0.583 & 31 & 0.845 & 31 & 0.856 \\ \hline
0.605 & 49 & 1.510 & 50 & 1.049 \\ \hline
0.627 & 74 & 3.004 & 67 & 2.905 \\ \hline
0.649 & 117 & 5.857 & 133 & 4.218 \\ \hline
0.671 & 173 & 4.144 & 186 & 5.285 \\ \hline
0.693 & 295 & 5.156 & 241 & 4.857 \\ \hline
0.715 & 391 & 6.293 & 351 & 7.604 \\ \hline
0.737 & 454 & 7.527 & 470 & 6.384 \\ \hline
0.759 & 573 & 3.400 & 551 & 2.840 \\ \hline
0.781 & 623 & 1.755 & 557 & 1.385 \\ \hline
0.803 & 515 & 1.173 & 513 & 1.660 \\ \hline
0.825 & 434 & 1.384 & 436 & 1.126 \\ \hline
0.847 & 365 & 0.824 & 370 & 1.380 \\ \hline
0.869 & 311 & 3.445 & 306 & 2.423 \\ \hline
0.891 & 219 & 1.281 & 258 & 0.589 \\ \hline
0.913 & 222 & 2.970 & 218 & 0.906 \\ \hline
0.935 & 179 & 1.010 & 196 & 0.654 \\ \hline
0.957 & 167 & 1.730 & 185 & 0.947 \\ \hline
0.979 & 159 & 9.783 & 137 & 0.754 \\ \hline
1.001 & 144 & 1.681 & 158 & 2.229 \\ \hline
1.023 & 104 & 0.581 & 119 & 0.848 \\ \hline
1.045 & 123 & 0.956 & 114 & 0.432 \\ \hline
1.067 & 118 & 0.568 & 99 & 0.910 \\ \hline
1.089 & 104 & 1.647 & 95 & 1.865 \\ \hline
1.111 & 88 & 0.645 & 88 & 0.532 \\ \hline
1.133 & 85 & 1.547 & 85 & 2.173 \\ \hline
1.155 & 75 & 2.167 & 85 & 1.934 \\ \hline
1.177 & 86 & 1.512 & 86 & 0.650 \\ \hline
1.199 & 61 & 0.668 & 67 & 1.473 \\ \hline
1.221 & 53 & 1.696 & 78 & 0.920 \\ \hline
1.243 & 60 & 0.480 & 64 & 1.038 \\ \hline
1.265 & 53 & 0.480 & 53 & 1.717 \\ \hline
1.287 & 51 & 0.844 & 47 & 0.626 \\ \hline
1.309 & 45 & 0.504 & 45 & 0.367 \\ \hline
1.331 & 36 & 0.432 & 41 & 1.835 \\ \hline
1.353 & 33 & 0.456 & 36 & 0.476 \\ \hline
1.375 & 30 & 0.371 & 39 & 0.928 \\ \hline
1.397 & 28 & 0.367 & 31 & 0.377 \\ \hline
1.419 & 15 & 0.788 & 24 & 0.295 \\ \hline
1.441 & 19 & 0.418 & 23 & 0.314 \\ \hline
1.463 & 18 & 0.417 & 8 & 0.390 \\ \hline
1.485 & 12 & 0.311 & 21 & 0.343 \\ \hline
1.507 & 15 & 0.336 & 21 & 0.324 \\ \hline
1.529 & 9 & 0.303 & 14 & 0.300 \\ \hline
1.551 & 9 & 0.251 & 14 & 1.428 \\ \hline
1.573 & 9 & 0.379 & 10 & 0.199 \\ \hline
1.595 & 10 & 1.623 & 11 & 0.141 \\ \hline
1.617 & 14 & 0.288 & 10 & 0.197 \\ \hline
1.639 & 7 & 1.113 & 5 & 0.172 \\ \hline
1.661 & 6 & 0.635 & 2 & 0.173 \\ \hline
1.683 & 5 & 0.163 & 4 & 0.173 \\ \hline
1.705 & 4 & 0.137 & 6 & 0.050 \\ \hline
1.727 & 5 & 0.012 & 3 & 0.100 \\ \hline
1.749 & 6 & 0.027  &  &  \\ \hline
1.771 & 1 & 0.330  &  &  \\ \hline
    \end{tabular}
\end{center}
\end{table}

\end{document}